\documentstyle[epsf,aps]{revtex}
\def\CA{{\cal A}}
\def\Journal#1#2#3#4{{#1} {\bf #2}, #3 (#4)}

\def\NPB{{\em Nucl. Phys.} B}
\def\NPA{{\em Nucl. Phys.} A}
\def\PLB{{\em Phys. Lett.} B}
\def\PRL{\em Phys. Rev. Lett.}
\def\PRD{{\em Phys. Rev.} D}
\def\PRC{{\em Phys. Rev.} C}

\begin{document}

\preprint{\vbox {\hspace*{\fill}DOE/ER/40762-155\\
          \hspace*{\fill}UMD PP\# 99-007}} 
\vspace{.5in}

\title {Systematic Power Counting in Cutoff Effective Field Theories for 
Nucleon-Nucleon Interactions and the Equivalence With PDS}

\author{Thomas D. Cohen}

\address{Department of 
Physics, University of~Maryland, College~Park, MD~20742-4111}

\author{James M. Hansen}

\address{Montgomery Blair  High School,  Silver Spring, MD~20901}

\maketitle

\vspace{.25in}

\begin{abstract}

An analytic expression for the ${}^1S_0$ phase shifts in
nucleon-nucleon scattering is derived in the context of the
Schr\"odinger equation in configuration space with a short distance
cutoff and with a consistent power counting scheme including pionic
effects. The scheme treats the pion mass and the inverse scattering
length over the intrinsic short distance scale as small parameters.
Working at next-to-leading order in this scheme, we show that the
expression obtained is identical to one obtained using the recently
introduced PDS approach which is based on dimensional regularization
with a novel subtraction scheme.  This strongly supports the
conjecture that the schemes are equivalent provided one works to the
same order in the power counting.
\vspace*{.15in}\end{abstract}

\vspace{.35in}

Weinberg \cite{Weinberg1} pointed out that there is a problem with using the
standard techniques of chiral perturbation theory ($\chi$PT) in nuclear physics
(as opposed to hadronic physics).  The difficulty is that nuclear physics has
low energy scales  quite distinct from $m_\pi$.  In low energy s-wave
scattering, the  scattering length, $a$,  is extraordinarily long on hadronic
scales ($\sim$ 23. fm in the ${}^1S_0$ channel and $\sim$ 6 fm in the ${}^3S_1$
\cite{scat} where $m_{\rho}^{-1} \sim .25$ fms).  These unnaturally long scales
set the scale at which a purely perturbative treatment breaks down.  This
suggests that a conventional perturbative approach breaks down at 
an absurdly low
momentum, $k \sim 1/a \approx$ 10 MeV in the ${}^1S_0$ channel.   Moreover,
$1/a$ plays the   role of the short distance scale (which we denote as 
$\Lambda$) in the perturbative problem so $m_\pi$ 
expansion will not converge.

There has been considerable effort  dealing with this ``unnatural  scattering
length'' problem during the past several years
\cite{Weinberg1,KoMany,Parka,KSWa,CoKoM,DBK,cohena,Fria,Sa96,LMa,GPLa,Adhik,RBMa,Bvk,aleph,Parkb,Gegelia,steelea,KSW,KSW2,CGSS}. 
The goal is to implement the general underlying philosophy of effective field
theory (EFT) \cite{EFT} to a problem in which $m_\pi$ is not the only light
scale.   Weinberg's proposal\cite{Weinberg1} was to use $\chi$PT to develop a
potential (two-particle irreducible graphs) which  would be iterated to all
orders yielding a Schr\"odinger  equation.    In order to implement chiral 
power counting for the
two-particle irreducible graphs,  four-nucleon  contact interactions are
required.  However, as noted by a number of authors\cite{KSWa,cohena,LMa} when
iterated to all orders, there is no straightforward way  to implement the
nonperturbative renormalization in a fashion consistent with the power
counting.  

Two main approaches have  emerged to get around this problem.  The first is
via cutoff EFT \cite{KoMany,Parka,CoKoM,cohena,GPLa,steelea}.  The 
scheme is to  put in 
{\it ad hoc} short-ranged form factors in front of the contact interactions.
By fitting the coefficients to get certain low energy observables (such as the
scattering length) correctly  one  renormalizes the interactions
in a way which is insensitive to the short distance physics. 
One can test this by  demonstrating explicitly the insensitivity of observables
to the cutoff scale\cite{Parkb}.  The scheme, as implemented to date, has 
the disadvantage,
however, that the calculation of observables has not 
been formulated directly in terms of an expansion parameter.  This obscures 
one of the main advantages of EFTs, the ability to make {\it a
priori} estimates of accuracy.

The second approach is the so-called pole divergent subtraction (PDS) scheme
of Kaplan, Savage and Wise\cite{KSW}.   The scheme is based on an effective
Lagrangian containing nucleons and pions and including nucleon contact
interactions.    In PDS the 
divergent integrals which arise from these interactions  are treated
via dimensional regularization with an unusual 
subtraction scheme: one subtracts poles at $\epsilon = 1$ from the 
analytically continued integrals rather than only  poles at $\epsilon =0$ as 
in the usual $\overline{MS}$
scheme.  While this prescription seems rather arbitrary, 
the consequence of this scheme is that 
coefficients in the Lagrangian have a systematic power counting
which is not destroyed by large values of $a$, and that  observables can be 
systematically expanded.  
A renormalization group analysis of the coefficients within this scheme 
shows that the approach requires that the lowest order s-wave interaction 
needs to
be iterated to all orders---building in a nonperturbative treatment suitable
for large scattering lengths---while all higher terms and all pionic effects
are treated perturbatively.  
The principal advantage of this scheme is quite clear,
one ends up with a treatment which is systematic in $Q/\Lambda$ 
where $Q \sim k,
m_\pi$ and thus in principle one can make {\it a priori} estimates of
errors due to higher order effects.  A number of observables have been
calculated with this approach at next-to-leading order including 
nucleon-nucleon scattering \cite{KSW},
and deuteron electromagnetic form factors \cite{KSW2} and 
polarizabilities\cite{CGSS}.

The purpose of this paper is  to show that systematic power counting rules for
observables can be implemented straightforwardly in the cutoff approach
and to demonstrate the equivalence of
the cutoff treatment with the PDS approach (provided both treatments implement
the same power counting rules and work to the same order). 
The explicit demonstration of this equivalence is useful, if for no other
reason, since it helps establish the validity of the formalisms.  
We explicitly demonstrate this equivalence in the context of 
scattering in the ${}^1S_0$ channel at next-to-leading order including the
pionic effects.
 
We implement a cutoff in configuration space by introducing a 
separation length scale, $R$, which essentially serves as a 
renormalization scale.  We parameterize the effects of
the physics at lengths  shorter than $R$ and calculate the effects for lengths 
greater than $R$.    The coefficients parameterizing the 
short distance physics vary
with $R$ in such a way as to offset changes in the long distance physics.  In
this way the formulation is renormalization group invariant.
As a practical matter, $R$ must be chosen to be sufficiently large
that the effects of the short distance physics are contained almost completely 
within $R$.  The  scale of the long distance physics is $m_\pi$
and we impose a  chiral expansion. Accordingly  we must assume formally 
that $m_\pi R \ll 1$.
  
We  work with a chiral expansion with $m_\pi a$ held fixed, which
allows for the unnaturally large value of $a$.  To simplify our power
counting, we  introduce a single low momentum scale, $Q$, and 
define the following power counting rules:
\begin{equation}
m_\pi \sim   Q  \; \; \; 
1/a \sim  Q \; \; \; 
k \sim  Q
\label{pc}
\end{equation}
where $k$ is the relative momentum of the colliding nucleons. This 
analysis based on power counting with a single low  scale is
consistent with the PDS analysis of ref.\cite{KSW}.  
A couple of innocent assumptions go into the derivation of the expression for
the phase shifts: i)  at large distances the nucleon-nucleon 
interaction can be described by a Schr\"odinger equation ; ii)  the 
energy dependence of the logarithmic derivative of the wave function
at $R$ is generically set by some short distance scale $\Lambda$.

 Weinberg's power counting\cite{Weinberg1} at the level
of the two-particle irreducible amplitudes is valid.  We  use the
Fourier transforms of these amplitudes into configuration space.   
The implementation of the $Q$ expansion is quite simple.  One works
in the Weinberg expansion up to some fixed order, including only the long
distance pion exchange terms which we denote as $V_{\rm long}$.
By hypothesis  $R$ is taken to be sufficiently large that the effects
of the short distance  potential are essentially contained within $R$ with an
accuracy comparable to or better than the order to which we are working.  
Thus by assumption, for $r>R$, the entire potential is given 
by $V_{\rm long}$. The associated Schr\"odinger equation is subject 
to boundary conditions at $R$,  which come from fixing the energy 
dependence of the short distance physics. 

For simplicity, we   consider the singlet channel and calculate $k
\cot (\delta)$ at next-to-leading order---which is ${\cal O}(Q^2)$.  The 
long range potential contributing at lowest nontrivial order is simply 
the central one-pion exchange
potential which is given in configuration space by
\begin{equation}
V_{\rm long}(r) \, = \,  V_{\rm opep}(r)\, = \, \frac{-g_A^2   m_\pi^2}{16 \pi f_\pi^2 } \, \frac{e^{-m_\pi r}}{r}
\label{opep}
\end{equation}
Note in this work we are using the convention that $f_\pi \approx$ 93 MeV. 
Thus, the $f_\pi$ used in this work is $1/\sqrt{2}$ times $f_\pi$ used in 
ref. \cite{KSW}.  For $r>R$ the system is described by the following radial 
Schr\"odinger
equation for the relative position.
\begin{equation} 
\left (-\frac{\partial^2}{\partial r^2} \, +  \, M \, V_{\rm long}(r)  - k^2
\right ) \,  u(r)
\, = \, 0
\end{equation}  
where the s-wave component of the wave function is given by $\psi_{s} = u/r$, 
and $M$ is the mass of the nucleon.
 
It is to useful express $u(r)$ as an expansion $
u(r) \, = \, u_0(r) \, + \, u_1(r) \, + \, u_2(r) + ...$
with $u_0(r)$ a solution of the noninteracting Schr\"odinger equation
and 
\begin{equation}
\left (-\frac{\partial^2}{\partial r^2} \,  - k^2 \right ) u_{i+1}(r) \, = 
 - \, M \, V_{\rm long}(r) \, u_{i} (r)
 \label{uip1} \end{equation}
It is immediately apparent that if the series for $u$ converges,
then any set of $u_i$'s which satisfy eq.~(\ref{uip1}) sums to a $u$ which
solves the original Schr\"odinger equation.  
The expansion is essentially a Born series  except that it
 is only valid for $r>R$ and has nontrivial boundary  conditions at $r=R$.
 We impose $u_i(R) = 0$ for $i \ge 1$ which puts all
 short distance physics effects into $u_0$.  
 Since $u_0$ is a solution of the free Schr\"odinger equation it can be
written as
$u_0(r) \, = \, \sin( k r + \delta_0 ) $;
 $\delta_0$ simply represents what the singlet phase shift would have been
had the long distance potential been removed.   

Simple dimensional analysis implies that each
term in the expansion for $u$ is order $m_\pi/\Lambda$ ($Q/\Lambda$) smaller 
than the previous term; one expects that as
$m_\pi \rightarrow 0$, the expansion for $u$ to converge. 
The leading nontrivial effect of the pions is contained in $u_1$.  It is 
straightforward to compute $u_1$ using Green's function methods---the
appropriate Green's
function  is 
 $\sin (k (r -r') )/k$ so that
\begin{equation}
u_1(r) \, = \,M \,  \int_R^r \, d r' \, \frac{\sin(k (r - r' ) )}{k} V_{\rm
long}(r')
u_0(r')
\label{u1}\end{equation} 
 We use the wave functions asymptotically in the determination of the phase
 shifts.
Inserting $u_0 = \sin(k r + \delta_0)$ into eq. (\ref{u1}) yields
\begin{equation}
\lim_{r \rightarrow \infty} u_1 (r) \, = \, \sin (k r) \, [
 \cos (\delta_0) f_1(k) \, + \, \sin(\delta_0)
f_2(k) ] \, - \, \cos(k r) \, [ \sin (\delta_0) \,  f_1 (k)  \, + \,   
\cos(\delta_0) \, f_3(k)]
 \end{equation}
 where
 \begin{eqnarray}
 f_1(k) \, & = & \, \frac{M}{k} \, \int_R^\infty \, d r' \,  V_{\rm long }(r') \sin (k
 r') \,  \cos ( k r') \\
  f_2(k) \, & = & \, \frac{M}{k} \, \int_R^\infty \, d r' \,  V_{\rm long}(r') \cos^2 ( k r') \\
 f_3(k) \, & = & \, \frac{M}{k} \, \int_R^\infty \, d r' \,  V_{\rm long}(r')   \sin^2 ( k r') 
 \label{fdef} \end{eqnarray}
Asymptotically the wave function is of the form $A \sin (k r + \delta)$ where 
A is the overall normalization. From this asymptotic form and the fact that to
this order $u = u_0 + u_1$ one can obtain an expression for $k \cot (\delta)$:
\begin{equation}
k \cot (\delta) \, = \, \frac{k \cot (\delta_0)\, (1 \, + \,  f_1) 
\, + \, k \,f_2  }{1 \,  - \, f_1 \, - \,  k \cot (\delta_0) \, (f_3 / k) }
\label{kcotd}
\end{equation}

We need to establish how the phase shift 
relation above behaves for small $Q$.  First, we note that we can
expand $k \cot (\delta_0)$ as a function of $k^2$.  This is an
effective range expansion for a system consisting only of the short distance 
part of the interaction:
\begin{equation} 
k \cot (\delta_0) \, = \, -1/a_{\rm short} \, + \, 1/2 \,r_e^0\,k^2 \, + \, v_2^0 \,  k^4 +
\, v_3^0 \, k^6 + \,v_4^0 \, k^8 + \ldots \label{effrange0}
\end{equation} 
By hypothesis,  the scale of the coefficients are
generically  set by the short distance scale $\Lambda$: $r_e^0 \sim
\Lambda^{-1}$,
$v_i^0 \sim \Lambda^{-2 i + 1}$.  This reflects
the assumption that at the matching distance, $R$, the energy dependence of the
logarithmic derivative of the wave function is set by $\Lambda$.  
Generically, one would  expect $1/a_{\rm short}\sim
\Lambda$ but here we will allow for the possibility that $1/a_{\rm short}$ 
is unnaturally small and will assign it a power counting scale of $Q$. As 
we will see subsequently  $1/a =1/a_{\rm short} + {\cal O}(Q^2/\Lambda)$.  
We are working to
order $Q^2$ in $k \cot (\delta)$ so we  truncate the expansion at the effective
range term.  It is useful to rewrite $1/a_{\rm short}$ in terms of its value in
the chiral limit plus a chiral correction
\begin{equation}
\frac{1}{a_{\rm short}} \,= \, \frac{1}{a_0} \, + \, d \,m_\pi^2
\end{equation}
where we have truncated the chiral expansion at order $m_\pi^2$.

Nontrivial chiral behavior is contained in the $f$ functions. 
Up to an overall multiplicative factor of   $(g_A^2   M)/(16 \pi f_\pi^2)$
the three $f$'s are functions $k$ with parameters  $m_\pi$ and $R$ and mass
dimension zero. 
One can write such functions as 
\begin{equation} 
f_i (k; m_\pi, R) \, \sim \, m_\pi\, \frac{-g_A^2   M}{16 \pi f_\pi^2 } \,  
\tilde{f}_i(k/m_\pi; m_\pi R)
\label{form}
\end{equation}
The preceding form is convenient in that we can do $Q$ 
power counting treating
$m_\pi R$ as a small parameter of order $Q/\Lambda$. Accordingly we
expand in   $m_\pi R$;  the lowest order
contributions come from taking  $m_\pi R =0$. 
Given a fixed $m_\pi$, this amounts to
taking $R$ to zero in the integrals.  For $f_1$ and $f_3$ this can be done 
quite straightforwardly:
\begin{eqnarray}
 f_1(k) \, & = & \, \frac{M}{k} \, \int_0^\infty \, d r' \,  V_{\rm long}(r') \sin (k
 r') \,  \cos ( k r') \, \left(1 + {\cal O}(m_\pi R)\right ) \nonumber\\
 & =  &\, -\,\, m_\pi 
 \frac{g_A^2 \,M}{32 \, \pi  f_\pi^2} \left(\frac{m_\pi}{k}\right )
 \tan^{-1} \left ( \frac{k}{m_\pi}\right ) \,  
\left(1 + {\cal O}(m_\pi R)\right ) 
 \nonumber 
 \end{eqnarray}
 \begin{eqnarray}  
 f_3(k) \, & = & \, \frac{M}{k} \, \int_0^\infty \, d r' \,  V_{\rm long} (r')
  \sin^2 ( k r') \left(1 + {\cal O}(m_\pi R)\right ) \nonumber\\
   &= &\,  - m_\pi 
  \, \frac{g_A^2 \,M}{64 \, \pi  f_\pi^2} \left (\frac{m_\pi}{k} \right )\, 
  \ln \left (1 + \frac{4 k^2}{m_\pi^2} \right ) \, 
\left(1 + {\cal O}(m_\pi R)\right ) \label{f} \end{eqnarray}
The corrections due to finite $m_\pi R$ contribute to the final results
only at higher order in $Q$ and  will be dropped here.
In evaluating $f_2$ one finds a logarithmic divergence when taking $R$ to zero.
It is useful to rewrite the $\cos^2$ in $f_2$ as $1- \sin^2$ so as to
re-express $f_2$ as
\begin{eqnarray}
f_2(k) \,& = &\, - f_3(k) \, + \, 
 \frac{M}{k} \, \int_R^\infty \, d r' \,  V_{\rm long}(r') \nonumber \\
 & = & \, -f_3(k) \, + \, 
 m_\pi \,    \frac{g_A^2 \,M}{16 \, \pi  f_\pi^2} \left (\frac{m_\pi}{k} \right )
 \, \left ( \gamma + \ln (m_\pi R)
 \right ) \, \left(1 + {\cal O}(m_\pi R)\right )
\label{f2}\end{eqnarray} 
The logarithmic dependence on R in this expression  will ultimately be absorbed 
in the renormalization process---by renormalization group invariance
$d$ has a $\ln (R)$ dependence against which the dependence in $f_2$
cancels.  

We can now insert our expressions for $k \cot(\delta_0)$ and the $f_i$
functions into  eq.~(\ref{kcotd}) and expand the resulting expression in $Q$ up to
$Q^2$ to obtain:
\begin{eqnarray}
k \cot (\delta) & = &-\frac{1}{a_0} \, + \, m_\pi^2 \, 
 \frac{g_A^2  M}{16 \pi f_\pi^2} \, \left (d+ \gamma + \ln (m_\pi R) \right)
 \, + \, \frac{1}{2} \, r_e^0 \,k^2 
\, - \,  \frac{1}{a_0^2} \, \frac{g_A^2  M}{64 \pi f_\pi^2} \,
\left( \frac{m_\pi^2}{k^2} \right )
\ln \left
(1 + \frac{4 k^2}{m_\pi^2} \right ) \nonumber \\
 & + &  \, \frac{m_\pi}{a_0} \, \frac{g_A^2  M}{16 \pi f_\pi^2} \, 
 \left( \frac{m_\pi}{k} \right )\,
 \tan^{-1} \left ( \frac{2 k}{m_\pi} \right ) \,
 + \, m_\pi^2 \, \frac{g_A^2  M}{64 \pi f_\pi^2} \, 
 \ln \left (1 + \frac{4 k^2}{m_\pi^2} \right ) 
 \label {kcotd1}\end{eqnarray}
Note that from eq.~(\ref{kcotd1}) and the fact that 
 $\left . k cot(\delta) \right|_{k=0} \, = \, -1/a$ one finds that 
 \begin{equation}
 - \frac{1}{a} \, = \, - \frac{1}{a_0} + \, m_\pi^2 \, 
 \frac{g_A^2  M}{16 \pi f_\pi^2} \, \left ( d+ \gamma + \ln (m_\pi R) \right)
 \, + \, \frac{g_A^2  M}{16 \pi f_\pi^2} \, \left( 
 \frac{2 m_\pi}{a_0} \, - \,\frac{1}{a_0^2} \right )  \, =  \,  
 \frac{1}{a_0} + {\cal O}(Q^2/\Lambda)
 \end{equation}

We wish to compare our result with that of the PDS scheme of
 Kaplan, Savage and Wise \cite{KSW}.
We will not review the scheme in detail here since ref. \cite{KSW} does a good
job of explaining clearly the steps necessary to implement the scheme.  

We take the relevant expressions from ref. \cite{KSW} and  use the notation
of ref. \cite{KSW} (except we  convert to our normalization of  
 $f_\pi$).  In the PDS scheme, one
calculates the scattering amplitudes directly order by order in $Q$, and 
infers the phase shifts or $k \cot (\delta)$ through them.   By unitarity, the
general form of the amplitude is given by 
\begin{equation}
\CA =  {4\pi\over M}{1\over (k \cot (\delta)  - i p)}\  \label{Amp}
\end{equation} 
  In the ${}^1S_0$  channel, apart from pion exchange
there are three nucleon contact operators in the Lagrangian which 
contribute at this order---there is contact interaction with no 
derivatives, a chiral correction to it and
two derivative interaction, parameterized by (scale-dependent)
coefficients $C_0$, $D_2$ and $C_2$, respectively.  The amplitude 
at order $Q^{-1}$ is obtained by
iterating the $C_0$ contact term to all orders.
\begin{equation}
{\cal A}_{-1} \, = \, { -C_0\over \left[1  + {C_0 M\over 4\pi} (ik + \mu)\right]}
 =  -{4\pi\over M}{1\over (1/a_0 + i k)} 
\label{Am1}\end{equation} 
where the second form can be taken as a definition of $a_0$. 
This definition is consistent with the definition used in the cutoff approach;
both correspond  to the value of the nucleon-nucleon scattering length
in an artificial  world where the quark masses were strictly zero. It is also
useful at this point to observe that the PDS formalism requires that $1/a_0$
is formally treated as being of order $Q$ since ${\cal A}_{-1}$ is order $1/Q$.
At order
$Q^0$, the scattering amplitude is given as the sum of five terms: 
${\cal A}_0={\cal A}_0^{(I)}+
{\cal A}_0^{(II)}+{\cal A}_0^{(II)}+{\cal A}_0^{(IV)}+{\cal A}_0^{(V)}$.
They are given by
\begin{eqnarray}
{\cal A}_0^{(I)} \, & = & \,  
-C_2 \, k^2
\left[ {\CA_{-1}\over C_0  } \right]^2
\nonumber \\
 {\cal A}_0^{(II)} \, &=& \,  {g_A^2\over 4 f_\pi^2} \, \left[-1 + {m_\pi^2\over
4 k^2} \ln \left ( 1 + {4 k^2\over m_\pi^2}\right)\right] \nonumber \\
 {\cal A}_{0}^{(III)} \, &=& \, {g_A^2\over 2 f_\pi^2} \left( {m_\pi M{\cal A}_{-1}\over 4\pi}
\right) \Bigg( - {(\mu + i k)\over m_\pi}
+ {m_\pi\over {2 k}} \left[\tan^{-1} \left({2 k\over m_\pi}\right) + {i\over 2} \ln
\left(1+ {4 k^2\over m_\pi^2} \right)\right]\Bigg)
\nonumber \\
{\cal A}_0^{(IV)} \, &=& \, {g_A^2\over 4 f_\pi^2} \left({m_\pi M{\cal A}_{-1}\over
4\pi}\right)^2 \Bigg(-\left({\mu + i k\over m_\pi}\right)^2
+ \left[ i\tan^{-1} \left({2 k\over m_\pi}\right) - {1\over 2} \ln
\left({m_\pi^2 + 4 k^2\over\mu^2}\right) + 1\right]\Bigg)
\nonumber\\
{\cal A}_0^{(V)} \, &=& \, - D_2 \, m_\pi^2 
\left[ {\CA_{-1}\over C_0  }\right]^2
\end{eqnarray}
where $\mu$ is the renormalization scale.  Equating 1 over the scattering amplitude from
eq.~(\ref{Amp}) with $1/(\CA_{-1} + \CA_0)$ and expanding to order $Q^2$ gives\cite{KSW}
\begin{equation}
k \cot(\delta) \, = \, i k \,  + \, \frac{4 \pi}{M \CA_{-1}} \,  - \,
\frac{ 4 \pi  \CA_0}{M \CA_{-1}^2} \label{ain} 
\end{equation}

Inserting the expressions for $\CA_{-1}$ and
$\CA_0$ into eq.~(\ref{ain}) and simplifying the resulting 
 expression  yields:
\begin{eqnarray}
k \cot (\delta) \, & = &\, -\frac{1}{a_0} \, + \, 
\left [ \frac{g_A^2  M}{16 \pi f_\pi^2} \, \left ( m_\pi^2 \,   \ln
\left (\frac{m_\pi}{\mu} \right ) \, - \,
 m_\pi^2 \, + \, \frac{1}{a_0^2} \, - \, 2 \frac{\mu}{a_0}  \, + \mu^2 \right )
\, + \, \frac{D_2 M}{4 \pi} \left ( m_\pi^2 \mu^2 \, -  \, 
\frac{2 m_\pi^2 \mu}{a_0} \, + \, \frac{m_\pi^2}{a_0^2} \right)  \right ] 
\nonumber \\ \nonumber\\
 \, & + & \, k^2  \, \left \{\frac{C_2 M}{4 \pi} \, \left ( \mu^2 \, - \, 
 \frac{2 \mu}{a_0} \, + \, \frac{1}{a_0^2} \right ) \right \} 
\, -  \,  \frac{1}{a_0^2} \, \frac{g_A^2  M}{64 \pi f_\pi^2} \,
\left( \frac{m_\pi^2}{k^2} \right ) 
\ln \left
(1 + \frac{4 k^2}{m_\pi^2} \right ) \nonumber \\ \nonumber\\
 \, & +  & \, \frac{m_\pi}{a_0} \, \frac{g_A^2  M}{16 \pi f_\pi^2} \, 
 \left( \frac{m_\pi}{k} \right )\,
 \tan^{-1} \left ( \frac{2 k}{m_\pi} \right ) \,
 + \, m_\pi^2 \, \frac{g_A^2  M}{64 \pi f_\pi^2} \, 
 \ln \left (1 + \frac{4 k^2}{m_\pi^2} \right ) 
 \label {kcotdpds}\end{eqnarray}
where to aid comparison with the eq.~(\ref{kcotd1}) we have used
eq.~(\ref{Am1}) to eliminate $C_0$
in favor of $a_0$.  Comparing eq.~(\ref{kcotd1}) with eq.~(\ref{kcotdpds})
we see the two are identical provided that we identify 
the quantities  in square
and curly brackets in  eq.~(\ref{kcotd1}) with  $m_\pi^2 \, 
(g_A^2  M)/(16 \pi f_\pi^2) \, \left (d+ \gamma + \ln (m_\pi R) \right)$ and 
$r_e^0/2$, respectively.  The quantities in square and curly brackets 
in eq.~(\ref{kcotdpds}) contain  coefficients from the Lagrangian which 
may be adjusted to reproduce the correct short distance physics and 
hence may be matched to the analogous terms in eq.~(\ref{kcotdpds}).  
This completes the demonstration that  expressions for ${}^1S_0$ 
scattering derived using the Schr\"odinger equation in configuration 
space is equivalent to the PDS result.
 
   The analysis based on the configuration space Schr\"odinger equation
is  along the line of conventional potential model treatments.  We 
simply imposed a power counting scheme in
$Q$ on the traditional approach.  The PDS approach is superficially very
different.  Our demonstration of the  equivalence confirms the 
supposition that  the PDS 
approach describes  the same physics provided one is in a regime in which 
the power counting in $Q$ is valid.  The extent to which we are in such a
regime will be addressed in a subsequent paper.
 
The authors thank Silas Beane and Daniel Phillips for numerous interesting
discussions. TDC gratefully acknowledges the support of the U.S. Department of
Energy under grant no. DE-FG02-93ER-40762.

\end{document}